\documentstyle[twoside,fleqn,espcrc2]{article}

\newcommand{\be}{\begin{equation}}
\newcommand{\ee}{\end{equation}}
\newcommand{\ba}{\begin{eqnarray}}
\newcommand{\ea}{\end{eqnarray}}

\newcommand{\AmS}{{\protect\the\textfont2
  A\kern-.1667em\lower.5ex\hbox{M}\kern-.125emS}}

\title{Quenched Chiral Perturbation Theory to one loop\thanks{Talk given by
    E. Pallante}} 

\author{G. Colangelo\address{ INFN - Laboratori Nazionali di Frascati, \\
P.O. Box 13, I-00044 Frascati (Rome), Italy }%
 \thanks{supported by the HCM, EEC--Contract No. CHRX--CT920026 
(EURODA$\Phi$NE).  }
        and 
     E. Pallante\address{Institut f\"ur Theoretische Physik, 
Universit\"at Bern, Sidlerstrasse 5,\\
CH--3012 Bern, Switzerland }%
\thanks{supported by Schweizerisches Nationalfonds} }
       
\begin{document}

\begin{abstract}
The divergences of the generating functional of quenched Chiral Perturbation
theory (qCHPT) to one loop are computed in closed form. We show how the
quenched chiral logarithms can be reabsorbed in the renormalization of the
$B_0$ parameter of the leading order Lagrangian. Finally, we do the chiral
power counting for the quenched case and show that also finite loop
corrections may diverge in the chiral limit. 
\end{abstract}

\maketitle

\section{Introduction}
Quenched CHPT has been recently formulated \cite{qCHPT} in order to reproduce
in a systematic way the quenched approximation extensively used on the
lattice. It is especially useful for estimating the size of quenching errors
in the finite volume case and for the better understanding of the sicknesses
(i.e. the absence of the chiral limit) caused by the anomalous behaviour of
the $\eta^\prime$ sector in the quenched approximation.
Our aim is to perform a complete renormalization of qCHPT at one--loop, on the
same line as what has been done by Gasser and Leutwyler in the ordinary CHPT
case \cite{gl}. We focus here on the calculation of the divergent part of the
generating functional to one--loop; the calculation is done for a generic
number of flavours $N$. Besides the divergent part, the finite part of the
one--loop corrections may diverge in the chiral limit. We illustrate this
behaviour by means of chiral power counting. In what follows
we work in Minkowski space-time, 
since our purpose here is to show how the quenched
approximation distorts the physical matrix elements, rather than comparing
with lattice results.

\section{One--loop divergences}

The leading order Lagrangian of qCHPT can be written as the graded-symmetry
generalization of the CHPT Lagrangian: ${\cal L}_2=$
\ba
&&\hspace{-.7cm}V_1(\Phi_0) \mbox{str} (D_\mu U_sD^\mu U_s^\dagger )
+V_2(\Phi_0) 
\mbox{str} (\chi^\dagger_s U_s+U_s^\dagger\chi_s )\nonumber\\
&&\hspace{-.7cm}-V_0(\Phi_0) 
+V_5(\Phi_0) D_\mu\Phi_0 D^\mu \Phi_0  ,
\label{L2}
\ea
where str is the supertrace and ${\cal L}_2$ is invariant under 
$SU_L(N|N)\otimes SU_R(N|N)\odot U(1|1)_V$. The graded meson field is 
$U_s = \exp (\sqrt{2} i\, \Phi/F)$, where $\Phi$ is a hermitian non traceless
$2\times 2$ block matrix
\[
\Phi = \left( 
\begin{array}{cc}
\phi & \theta^\dagger  \\
\theta & \tilde\phi
\end{array}  \right) \; , ~~~\mbox{str}(\Phi) = \Phi_0 =
\phi_0-\tilde\phi_0 \; \; ,
\]
which contains the {\em ghost} states of the quenched spectrum: $ \tilde\phi$
is a bosonic state with the quantum numbers of a 
$\tilde{q}\bar{\tilde{q}}$ pair, 
 $\theta$ and $\theta^\dagger$ are the fermionic hybrid states 
$\tilde{q}\bar{q}$ and $q\bar{\tilde{q}}$ respectively. The graded external
sources are introduced in the usual manner:
$D^\mu U_s = \partial^\mu U_s - i r_s^\mu U_s+iU_s l_s^\mu$,
$\chi_s=2B_0(s_s+ip_s)$, with the scalar source 
$s_s={\mbox{diag}} ({\cal M},\;{\cal
  M})+\delta s_s$ and we work in the degenerate mass case 
${\cal M} = m_q { \bf 1}$. Since we are only interested in the physical matrix
elements, we set to zero all the spurious external sources coupled to the
ghost fields. Finally, the potentials $V_i(\Phi_0)$ are even and real
functions of the singlet field $\Phi_0$ and we expand them as follows (we use
$N_c =3$):
\ba
V_0(\Phi_0) &=& \frac{m_0^2}{2N_c} \Phi_0^2 +O(\Phi_0^4) \nonumber \\
V_{1,2}(\Phi_0) &=& \frac{F^2}{4}+{1\over 2}v_{1,2}\,\Phi_0^2 + O(\Phi_0^4)
\nonumber \\
V_5(\Phi_0) &=& \frac{\alpha}{2N_c} + O(\Phi_0^2) .  
\ea

The divergences of the quenched generating functional (GF) 
to one loop are obtained
with the background field method, i.e. expand the leading order
action around the
classical solution of the eqs. of motion. We write the field 
$U_s= u_s ~e^{i \Xi}~ u_s$, where $\bar{U_s}=u_s^2 $ is the classical solution
and $\Xi$ is the fluctuation matrix
\[
\Xi = \left( 
\begin{array}{cc}
\xi & \zeta^\dagger  \\
\zeta & \tilde\xi
\end{array}  \right) \;\; , \; \; \; \; \mbox{str}(\Xi)=
\sqrt{N}(\xi_0-\tilde\xi_0) .
\]
The relevant action quadratic in
the fluctuations can be written in a diagonal form: $S[\Phi]=S[\bar{\Phi}]$
\be
 - \frac{F^2}{4} \int dx \left\{  \xi^T D_\xi \xi
+2 \zeta^\dagger D_\zeta \zeta+ X_0^T {\bar{D}}_X X_0\right\} ,
\label{S}
\ee
where $D_\xi ,D_\zeta$ carry flavour indices, 
$\xi^T =(\xi^1, \xi^2,$ $\ldots, \xi^{N^2-1})$, 
$\zeta^\dagger =(\zeta^{\dagger 0}, \zeta^{\dagger 1}, \ldots, 
\zeta^{\dagger N^2-1})$ and $X_0^T =(\xi_0, \tilde\xi_0)$. Notice that 
the three quadratic forms in eq. (\ref{S}) are the standard CHPT form, 
the fermionic ghost form and the pure singlet form respectively. 
The singlet operator $\bar{D}_X$ acts
on the graded group and is given by:
\ba
\bar{D}_X &=& D_X^0+A_X -{1\over 2}(1+\tau_3) B^T D_\xi^{-1} B \nonumber\\
D_X^0&=& \tau_3(\Box +M^2)+\frac{N}{3}(1-\tau_1)(\alpha \Box +m_0^2)
\nonumber \\
A_X &=& \frac{1}{4N}(1+\tau_3) \langle \hat\chi_+ \rangle
 -N(1-\tau_1)\left ( v_1 \langle u_\mu u^\mu \rangle \right .\nonumber\\
&&\left . +v_2 \langle
\hat\chi_+ \rangle \right ) +O(\Phi_0^2) \nonumber \\
B^a &=& \frac{1}{2\sqrt{2N}} \langle \lambda^a \chi_+ \rangle , 
\label{DX}
\ea 
where  $\langle\hat\chi_+\rangle =\langle\chi_+ \rangle -2NM^2$, 
$M^2=2B_0 m_q$ and $u_\mu = iu^\dagger D_\mu U u^\dagger$. 
From eq. (\ref{S}) the GF to one loop can be formally written as
\be
e^{iZ^{\mbox{\tiny{qCHPT}}}_{\mbox{\tiny{1~loop}}}} = 
{\cal N} {\det D_\zeta\over (\det D_\xi )^{1\over 2}(\det
  \bar{D}_X)^{1\over 2}} .
\label{ZQ}
\ee
The $\ln\det D_\xi$ and $\ln\det D_\zeta$ are calculated with the standard heat
kernel technique, while in the singlet case we perform a perturbative expansion
of $\mbox{Tr} \ln \left(\overline{D}_X/D_X^0 \right)$ around the non diagonal
kinetic term $D_X^0$. Despite the non locality of $\bar{D}_X $ the 
divergent part of the
quenched GF admits a closed form. The result reads:
$Z^{\mbox{\tiny{qCHPT}}}_{\mbox{\tiny{1~loop}}}=$ 
\ba 
&&\hspace{-.7cm}-\frac{1}{(4\pi)^2(d\!-\!4)} 
\int dx \left[
{1 \over 8} \langle u_\mu u_\nu \rangle \langle u^\mu u^\nu \rangle +
{1 \over 16} \langle u_\mu u^\mu \rangle^2 \right. \nonumber \\
&&\hspace{-.7cm}+ {1\over 8} \left(1 - 4 v_1 \right) \langle  u_\mu
u^\mu \rangle \langle \hat\chi_+ \rangle 
 + {1\over 16} \left(1 - 8 v_2 \right)\langle \hat\chi_+
 \rangle^2 \nonumber \\ 
&&\hspace{-.7cm}+{m_0^2\over 6}\langle \chi_+ \rangle
 +{\alpha^2\over 72} \langle \hat\chi_+ \rangle^2
-{\alpha\over 12} \langle \chi_+^2 \rangle 
\nonumber\\
&&\hspace{-.7cm}\left.- {1\over 4}
\langle  u_\mu \rangle \langle u^\mu \left(u_\nu
    u^\nu + \chi_+\right) \rangle  
\right] +\ldots \, .
\label{ZZ}
\ea
We list the main properties of the divergent structure of
qCHPT to one--loop: 1) all the terms proportional to $N$ have been dropped by
means of the cancellation between $D_\xi$ and $D_\zeta$ determinants, 2)
 all the terms proportional to $1/N, 1/N^2$ have been dropped  by
means of the cancellation between $D_\xi$ and $\bar{D}_X$ determinants, 3)
new divergences proportional to the parameters of the singlet sector have been
produced. This recipe can be used to derive {\em a priori} the quenched GF for
weak interactions once the $N$ dependence of the fundamental one is known
\cite{long}.

\subsection{ Renormalization of $B_0$}

In eq. (\ref{ZZ}) only the term $m_0^2 \langle\chi_+\rangle$ gives rise to the
quenched chiral logs of the form $m_0^2\log M_\pi^2$, which diverge in the
chiral limit as it is well known. This term is a chiral invariant already
present in the leading order $(p^2)$ Lagrangian. To remove that divergence in
the renormalization procedure one has to add a divergent counterterm to the
lowest order parameter $B_0$. The renormalized one reads as follows:
\be
B_0 \rightarrow \overline{B}_0 = B_0 \left( 1- \frac{m_0^2}{48
   \pi^2 F^2}  \ln \frac{M^2}{\mu^2} +b_0(\mu)\right) ,
\label{B0bar}
\ee   
where $\mu$ is the renormalization scale. We stress that this feature is new
with 
respect to standard CHPT and stems from the fact that in the quenched case a
new mass scale $m_0$ is present and does not vanish in the chiral limit.
After renormalization, the pion mass is given by $M_\pi^2=2\bar{B}_0m_q
+O(m_q^2)$, which behaves well in the chiral limit:
\be
\lim_{m_q\to 0} M_\pi^2\sim \bar{B}_0m_q\sim m_q\ln m_q \to 0 .
\ee
Observables on the lattice can be divided into two categories: the ones
proportional to $M_\pi^2$ and the ones proportional to $\bar{B}_0$ only (they
are the $\bar{q}q$ matrix elements). The first observables cannot show the
divergence due to quenched chiral logs, once $M_\pi^2$ renormalized is used.
The second ones do show the divergence due to quenched chiral logs through the
 $\bar{B}_0$ parameter. Some examples are the scalar quark condensate 
$\langle \bar{q}q\rangle_q = -F_\pi \bar{B}_0 \left[1 + O(M^2)
\right]$ and the scalar form factor of the pion $F_S^q(t)=2
\bar{B}_0[1+O(t)]$. These are the candidates to disentangle quenched chiral
logs on the lattice.

\section{Finite loop corrections}

In the quenched case also the finite part of the one--loop corrections may
diverge in the chiral limit. These divergences are power-like and therefore
they can be dominant with respect to any logarithmic divergence. 
Their origin can
be traced back to the presence of the double pole in the singlet propagator as
it happens in the case of quenched chiral logs. The double pole carries the
new $m_0^2$ vertex which has chiral dimension zero. Thus, while in standard
CHPT the lowest chiral order of a vertex is two, in qCHPT this is zero.
The chiral order $D_g$ of a generic diagram with $L$ loops, $N_d$ vertices of
dimension $d$ is $D_g = 2 L + \sum_d (d-2) N_d + 2 $. In CHPT $d\geq 2$
implies $D_g\geq 2$, while in qCHPT $d=0$ implies that $D_g$ can be
negative. It follows the presence of power--like corrections of the type 
$1/(M_\pi^2)^{n\geq N}$ which diverge in the chiral limit. 
The lowest power $N$ 
allowed for a given observable is dictated by a set of rules in constructing
quenched diagrams: i) only one $m_0^2$ vertex can be inserted in each singlet
line, ii) no standard vertices can have all outgoing lines ending on an
$m_0^2$ vertex (the only exception being vertices with external sources).
An interesting example is the scalar radius of the pion. It turns out that the
quenched one--loop contribution diverges not just logarithmically as in
standard CHPT, but like an inverse power of the pion mass: 
\ba
&&\hspace{-.7cm}\langle r^2\rangle^q_S\vert_{M_\pi\to 0}\, \sim\,  
{1\over 16\pi^2 F_\pi^2}
\left [ ~{4\over 45}{m_0^4\over M_\pi^4}\, + \right . \nonumber\\
&&\hspace{-.7cm}\left . -{1\over 3}\left ( 1-{2\over 15}\alpha\right ) 
{m_0^2\over M_\pi^2} -3(1-4v_1) \log M_\pi^2\right ]\, .
\ea
$1/M_\pi^4$ is also the lowest chiral power which can contribute to $\langle
r^2\rangle^q_S$. The one--loop coefficients of the power--like terms may
receive corrections from higher loops, which are suppressed by extra powers of
$(m_0^2/N_c)/16\pi^2F_\pi^2$.
We also computed the full one--loop contribution to the $S$--wave scattering
lengths in the $I=0,2$ channels \cite{long}. 
A numerical estimate of the expected to be
dominant contributions at one--loop (the singlet $m_0^2$ insertions and the
standard chiral logs) is given in Table \ref{taba00} for the $I=0$ case 
as a function of the pion
mass and with $\mu =1$ GeV in the chiral logs. A similar trend in the $I=2$
channel is obtained. We defined  $\epsilon = 
{M_\pi^2 / 48\pi^2 F_\pi^2}$ and $\bar{\delta} =  {m_0^2-\alpha M_\pi^2
 /48\pi^2 F_\pi^2}$. At the physical value of the pion mass the 
$\bar{\delta}^2 / \epsilon$ term is largely dominant, i.e. quenching artifacts
are dominant. At larger values of the pion mass (the typical ones in a lattice
simulation) standard chiral logs start soon to be dominant and their
coefficient happens not to be substantially changed by quenching $(\leq 30\%)$
\cite{plb}.
\begin{table}
\caption{Numerical values of the leading contributions to $a_0^0$
  quenched up to one loop.}
\label{taba00}
\begin{center}
\begin{tabular}{ccccc}
\hline
$M_\pi$ (MeV) & tree & $\bar{\delta}$ &  $\bar{\delta}^2/\epsilon$ &
$\epsilon \ln M_\pi^2$ \\
\hline 
140 & 7 & 0.09 & 2.5 & 1.25 \\
300 & 7 & 0.08 & 0.47 & 3.5 \\
600 & 7 & 0.06 & 0.06 & 5.9 \\
\hline
\end{tabular}
\end{center}
\end{table}

\end{document}